\documentclass{iopart}

\usepackage{epsfig}
\usepackage{iopams}
\usepackage{amssymb}
\usepackage{bm}
\usepackage{color}

\newcommand{\beq}{\begin{equation}}
\newcommand{\eeq}{\end{equation}}
\newcommand{\beqarray}{\begin{eqnarray}}
\newcommand{\eeqarray}{\end{eqnarray}}

\newcommand{\eq}[1]{(\ref{#1})}  
\newcommand{\fig}[1]{figure~\ref{#1}}  
\newcommand{\Sec}[1]{section~\ref{#1}}  
\newcommand{\Ref}[1]{\cite{#1}}

\begin{document}

\title{Interface currents in topological superconductor-fer\-ro\-mag\-net
  heterostructures}
\author{P M R Brydon$^1$, Carsten Timm$^2$ and Andreas P Schnyder$^3$}
\address{$^1$ Institut f\"{u}r Theoretische Physik, Technische Universit\"{a}t  
  Dresden, D-01062 Dresden, Germany}
\ead{brydon@theory.phy.tu-dresden.de}
\address{$^2$ Institut f\"{u}r Theoretische Physik, Technische Universit\"{a}t  
  Dresden, D-01062 Dresden, Germany}
\ead{carsten.timm@tu-dresden.de}
\address{$^3$ Max-Planck-Institut f\"ur Festk\"orperforschung,
  Heisenbergstrasse 1, D-70569 Stuttgart, Germany}
\ead{a.schnyder@fkf.mpg.de}

\date{\today}

\begin{abstract}
We propose the existence of a substantial charge current  parallel to
the interface between a noncentrosymmetric superconductor and a metallic
ferromagnet. Our analysis focuses upon two complementary
orbital-angular-momentum pairing states of the superconductor, exemplifying
topologically nontrivial states which are gapped and gapless in
the bulk, respectively. Utilizing a quasiclassical scattering
theory, we derive an
expression for the interface current in terms of
Andreev reflection coefficients. Performing a systematic study of the
current, we find stark qualitative differences between the gapped and
gapless superconductors, which reflect the very different underlying
topological properties. For the
fully gapped superconductor, there is a sharp drop in the
zero-temperature current as the system is tuned from a topologically
nontrivial to a trivial phase. We explain this in terms of the
sudden disappearance of the contribution to the current from the
subgap edge states at the topological transition. 
The current in the gapless superconductor is characterized by a
dramatic enhancement at low temperatures, and exhibits a singular
dependence on the exchange-field strength in the ferromagnetic
metal at zero temperature. This is caused by the energy shift
of the strongly spin-polarized nondegenerate zero-energy flat
bands due to their coupling to the exchange field.
We argue that the interface current provides a novel test of the
topology of the superconductor, and discuss prospects for the
experimental verification of our predictions.
\end{abstract}


\maketitle

\section{Introduction} \label{sec:intro}

The discovery that gapped single-particle Hamiltonians can have a
nontrivial topology, depending on their dimensionality and the presence of
time-reversal and particle-hole
symmetries~\cite{schnyder2008,Kitaev2009,hasanKane2010,ryuNJP10,qiZhangReview2010},
has sparked a massive search for 
topological materials. A key motivation is to realize exotic Majorana-fermion
states, which are guaranteed to appear at the edges of a fully gapped 
topological insulator or superconductor by the bulk-boundary correspondence~\cite{ryuNJP10,qiZhangReview2010},
and which may have applications in quantum
computation. Parallel to these developments, the concept of topological
nontriviality has been 
generalized to \emph{gapless} systems, such as nodal
superconductors~\cite{satoPRB06,Beri2010,Volovik,yada2011,Schnyder2011,Schnyder2012,Dahlhaus2012,Matsuura2012} or Weyl
semimetals~\cite{Wan2011,Turner2013}. Bulk-boundary correspondences can also
be developed in these 
cases, leading to the topologically protected appearance of nondegenerate
zero-energy (or Majorana) arc lines or flat bands at certain surfaces. 
Much work has now been done on understanding the conditions under which these
states can form, and a topological 
classification of stable Fermi  surfaces of any dimension has recently
been developed~\cite{Matsuura2012,Zhao2012}.

\begin{figure}
\begin{center}
\includegraphics[clip,width=\columnwidth]{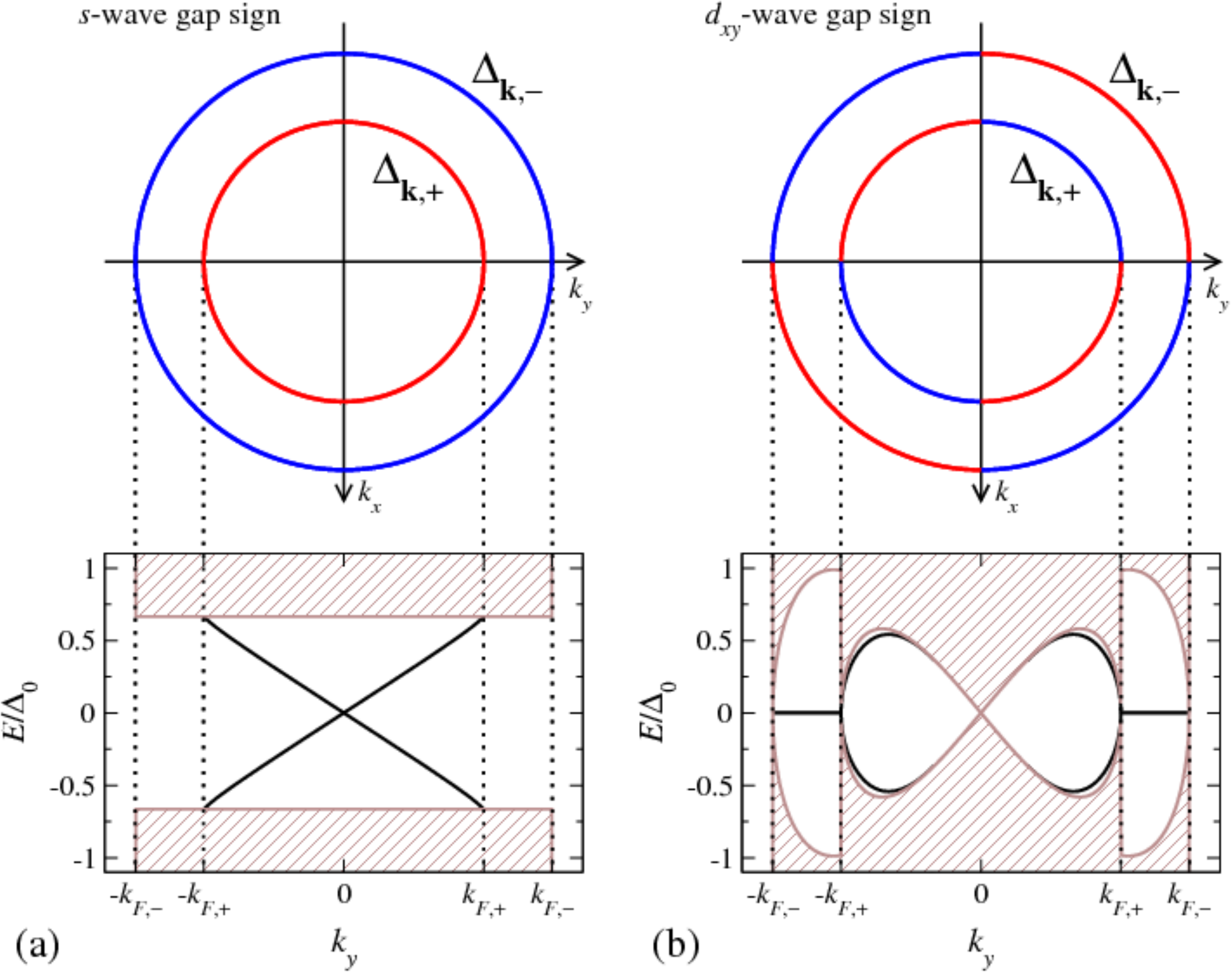}
\end{center}
  \caption{\label{projection} The relationship of the bulk gap structure to
    the (10) edge states of an NCS with
    (a) $s$-wave and (b) $d_{xy}$-wave form factors in a majority-triplet
    state. At the top of the
    figure we show the variation of the sign of the positive- and
    negative-helicity gaps, $\Delta_{{\bf 
        k},+}$ and $\Delta_{{\bf k},-}$ respectively, about the spin-orbit split
    Fermi surfaces. Here red (blue) indicates
    positive (negative) sign, and we assume that the positive-helicity
    (negative-helicity) Fermi  
    surface is circular with Fermi wavevector $k_{F,+}$ ($k_{F,-}$). Below we
    show the spectrum at the (10) edge for singlet-triplet parameter
    $q=0.25$, where black lines indicate edge states while the brown shaded
    region is the projection of the continuum onto the one-dimensional edge Brillouin
    zone. }   
\end{figure}

A promising class of materials in which to search for topological systems are
noncentrosymmetric superconductors (NCS). The absence of bulk
inversion symmetry in these compounds has two important consequences: it leads
to a strong  momentum-antisymmetric spin-orbit coupling and it permits the
existence of mixed-parity pairing states with both  
singlet and triplet gaps present~\cite{frigeriPRL04}. These exotic
superconducting properties have inspired a strong research
effort~\cite{bauerSigristbook}, and many examples of unconventional 
superconductivity in NCS have been reported, e.g., CePt$_3$Si~\cite{Bauer2004},
CeRhSi$_3$~\cite{Kimura2005}, CeIrSi$_3$~\cite{Sugitani2006},
Li$_2$Pt$_3$B~\cite{Li2Pt3B}, Y$_2$C$_3$~\cite{Chen2011} and
BiPd~\cite{mondal12}.  
More recently, much attention has been focused on the possible nontrivial
topology of
NCS~\cite{satoPRB06,yada2011,Schnyder2011,Schnyder2012,Dahlhaus2012,Matsuura2012,Qi2009,satoFujimoto2009,QiHughesZhangPRB2010,Yip2010,Tanaka2009,Tanaka2010,sato2011,Brydon2011}. Specifically,
the BCS Hamiltonian of an NCS belongs to class 
DIII of 
the Altland-Zirnbauer classification scheme. In two dimensions, gapped DIII
systems may be topologically nontrivial and possess a nonzero
$\mathbb{Z}_{2}$ topological number. An example of such a state is given by
the NCS with Rashba spin-orbit coupling, $s$-wave form factor  of the gap, and
majority-triplet 
pairing~\cite{Qi2009,satoFujimoto2009,QiHughesZhangPRB2010,Yip2010}. As shown
in~\fig{projection}(a), the edge spectrum 
possesses helically dispersing subgap states with Majorana zero
modes~\cite{satoFujimoto2009,Tanaka2009,Iniotakis2007,Vorontsov2008}, as
required by the bulk-boundary correspondence. In analogy to a
quantum spin Hall insulator, the edge states carry a spin
current~\cite{Tanaka2009,Vorontsov2008,LuYip2010}. 
Increasing the strength of the singlet pairing ultimately leads to a sign
change of the negative-helicity gap, which marks the transition into a state
with trivial topology~\cite{QiHughesZhangPRB2010}. The edge spectrum of
this state does not display any subgap states. 

It is likely that many NCS are characterized by rather strong interactions,
which may lead to higher orbital-angular-momentum pairing
states~\cite{yokoyama07}, e.g., the modulation of the gaps by a $d_{xy}$-wave
form factor as shown in~\fig{projection}(b).
Because of the presence of gap nodes, it is not possible to define a quantized
global topological number for such an NCS. This NCS nevertheless also displays
edge states with 
 nontrivial topology. Every point in the $(lm)$ edge Brillouin zone not
lying on a projected gap node may be regarded as the edge of a fully gapped
one-dimensional Hamiltonian in Altland-Zirnbauer class AIII. This allows the
definition of a momentum-dependent winding number $W_{(lm)}$, which can only
change across projected gap nodes~\cite{yada2011,Schnyder2011}. In
the case of 
the (10) edge shown in~\fig{projection}(b), the winding number $W_{(10)}$
evaluates to $\pm1$
for $k_y$ between the projected edges of the spin-orbit-split Fermi
surfaces. By the
bulk-boundary correspondence for the one-dimensional class AIII
Hamiltonian, it follows that there is a nondegenerate zero-energy flat 
band at these momenta~\cite{yada2011,Tanaka2010,sato2011}. The $d_{xy}$-wave
form factor is
crucial here, as one may show that these states can only form when there is a
sign difference between the gap on the forwards- and backwards-facing parts of
the Fermi surface~\cite{Schnyder2012,Brydon2011}. As such, these states are
present for 
both majority-triplet and majority-singlet pairing states. In contrast, the
edge states at momenta $k_y$ lying between the projected edges of the
positive-helicity Fermi surface depend upon the relative strength of the
singlet and 
triplet gap: for a majority-triplet state there are topologically trivial
dispersing states, whereas for majority-singlet pairing there are doubly
degenerate zero-energy states with $W_{(10)}=\pm2$.

Much attention has been paid to the spin structure of the edge states of  an 
NCS~\cite{Tanaka2009,Vorontsov2008,LuYip2010}, as the surface spin current
can be understood in terms of the polarization of the \emph{electronlike} part of
the edge state wavefunctions.
We have recently shown~\Ref{Schnyder2013} that the edge states of
an NCS typically also
have rather strong \emph{total} spin polarization, which 
consists of contributions from the electronlike and holelike components of the wavefunctions.
Like the  polarization of the electronlike wavefunction components \cite{Tanaka2009,Vorontsov2008,LuYip2010}, the total spin polarization is
odd in  
the edge momentum as required by time-reversal symmetry, and depends on both the
spin-orbit coupling and the relative 
strength of the 
singlet and triplet pairing. On the other hand, the total spin polarization
can be rather different from the electronlike spin
polarization, and it is the total polarization
that couples to an external exchange field. In particular, we have demonstrated
that the topologically protected
zero-energy 
flat bands characteristic of the NCS with $d_{xy}$-wave form 
factor have particularly strong total spin polarization. Coupling
to an exchange field therefore gives opposite energy shifts
to these flat bands on either side of the edge Brillouin zone, hence
generating an imbalance between  
the integrated spectral density at these two momenta. This causes the
appearance of a finite edge current, which depends rather strongly
on the orientation of the exchange field, and shows a remarkable singular
dependence on the exchange-field strength at zero temperature. 
In contrast, the absence of the flat bands for an NCS with $s$-wave
form factor leads to a very weak edge current due to the
interplay between the spin-orbit coupling and the spin
polarization induced by the exchange field.

The analysis in Ref.~\Ref{Schnyder2013} was performed for
an NCS strip in contact with a ferromagnetic \emph{insulator} at
one edge. The strip was described by a lattice model
with an exchange field applied to the sites on one
edge. Note that  applying an exchange field to the entire NCS
also leads to an
energy shift of the edge states~\cite{Matsuura2012,sato2011,WongLee2012} but
additionally distorts the Fermi surfaces in a way which is inconsistent with a
zero-momentum pairing state~\cite{AgtKau2007,Loder2012}.
In the present paper we consider an NCS
in proximity contact with a bulk \emph{metallic} ferromagnet (FM),
with each phase occupying a half-space and treated in the continuum limit.
We note that such
an NCS-FM heterostructure has been studied by other
authors~\cite{Duckheim2011,Annunziata}, but they only address the proximity
effect on the FM which does not concern us here. 
A fundamental difference between Ref.~\cite{Schnyder2013} and the present work
is that the direct coupling between the edge states and the exchange field in
the former is absent in the 
latter. While the edge states remain well-defined when an 
exchange field is applied to the edge layer~\cite{Schnyder2013}, for the
half-space continuum system
tunneling into the bulk states of the FM gives them a finite lifetime
and hence turns them into broadened resonances. One of our main
goals is to
understand how these differences affect the edge current. To
accomplish this, we utilize a quasiclassical technique to express the
interface current in terms of Andreev reflection processes.
The quasiclassical technique also has the advantage of being able
to treat realistically small superconducting gaps, whereas the
exact-diagonalization approach used in Ref.~\cite{Schnyder2013} requires rather
large gaps to avoid finite-size artifacts. We perform
a systematic study of the current's dependence on the temperature,
the pairing state of the NCS, and the exchange field  in the FM. For the  
$d_{xy}$-wave form factor, we find that the key features of the
interface current are robust to the additional complications of a
metallic FM. In contrast, the results for the $s$-wave form factor
show that the broadening of the subgap interface states leads to
qualitatively different behaviour of the current at low temperatures.
We use the close
relationship between the current and the interface local density of
states to understand the origin of the current, and show how it
reflects the topology of the NCS.

Our paper is organized as follows. We commence in~\Sec{sec:theory}
with the theoretical description of the system, including the construction of 
the scattering wavefunctions, an ansatz for the Green's function in the NCS
and the derivation of the current. In~\Sec{sec:results}, we present
and discuss the results for the $s$-wave and $d_{xy}$-wave gap form
factors. This is followed in~\Sec{sec:experiment} by a discussion of
possible experiments. We summarize our work
in~\Sec{sec:summary}.

\section{Theory} \label{sec:theory} 

\begin{figure}
\begin{center}
\includegraphics[clip,width=0.5\columnwidth]{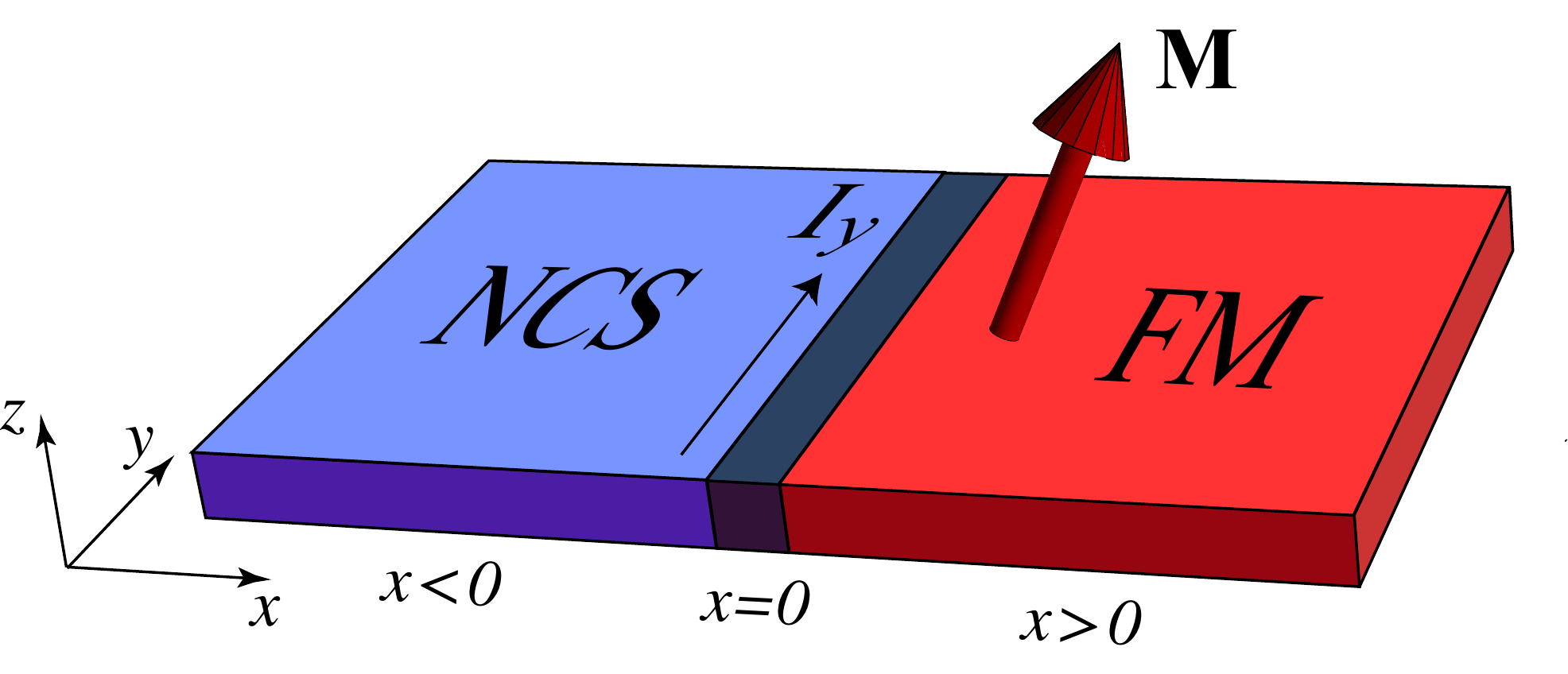}
\end{center}
  \caption{\label{heterostructure} Schematic diagram of the
    NCS-FM heterostructure considered in this work. The NCS and FM occupy the
    $x<0$ and $x>0$ half spaces, respectively, and are separated by 
an insulating barrier.  The magnetization ${\bf M}$ of the FM is allowed to
point in any direction. In general, we find that an interface current $I_{y}$
is present in the NCS as indicated by the black arrow.}
\end{figure}

We study a planar junction between a bulk NCS
and a metallic FM separated by a thin insulating barrier,
see~\fig{heterostructure}. The Hamiltonian is written as
\beq
\fl H = \int d^2r\, \underline{\Psi}^\dagger({\bf r})\check{H}({\bf
r})\underline{\Psi}({\bf r})  =  \int d^2r\, \underline{\Psi}^\dagger({\bf
r})\left(\begin{array}{cc}
\hat{H}_{0}({\bf r}) & \hat{\Delta}({\bf r}) \\
\hat{\Delta}^{\dagger}({\bf r}) & -\hat{H}^{T}_{0}({\bf r}) 
\end{array}\right)\underline{\Psi}({\bf r})\, \label{eq:Ham}
\eeq
where $\underline{\Psi}({\bf r})$ is the Nambu spinor of field operators. We
also introduce the notation of the hat and the caron (inverted hat) to
denote $2\times2$ matrices in spin space and $4\times4$ matrices in Nambu-spin
space, respectively. The noninteracting Hamiltonian is 
\beqarray
\fl \hat{H}_{0}({\bf r}) =  \left[-\frac{\hbar^2\pmb{\nabla}^2}{2m} - E_F +
  U\delta(x)\right]\hat{\sigma}_{0}   - \mu_{B}\Theta(x){\bf H}_{\rm{ex}}\cdot{\hat{\pmb\sigma}}\nonumber \\
 + \frac{\lambda}{2}\left\{\Theta(-x)\left[(-i\pmb{\nabla})\times{\bf
    e}_z\right]\cdot\hat{\pmb{\sigma}} + \left[(-i\pmb{\nabla})\times{\bf
    e}_z\right]\cdot\hat{\pmb{\sigma}}\Theta(-x)\right\} \,. \label{eq:Ham0}
\eeqarray
For simplicity, we assume that the effective mass $m$ and Fermi
energy $E_F$ are the same in the NCS and FM. The insulating layer
separating the NCS and FM is modeled as a $\delta$-function barrier of
strength $U$. In the NCS there is a Rashba spin-orbit coupling 
$\lambda$, while the FM is characterized by an exchange field ${\bf
  H}_{\rm{ex}} = |{\bf H}_{\rm{ex}}|\,(\cos\eta\sin\zeta\,{\bf e}_x +
\sin\eta\sin\zeta\,{\bf e}_y + \cos\zeta\,{\bf e}_z)$; note that the
spatially-dependent spin-orbit coupling in~\eq{eq:Ham0} is written in
symmetrized and thus  Hermitian form. The spin degeneracy of
the circular Fermi surfaces is 
therefore lifted in both the NCS and the FM: in the NCS we have
positive-helicity ($+$)
and negative-helicity ($-$) Fermi surfaces with radii
$k_{F,\pm} = k_F(\sqrt{1 + \widetilde{\lambda}^2} \mp \widetilde{\lambda})$
where $\widetilde{\lambda} = \lambda k_F/ ( 2E_F )$ and $k_{F} =
\sqrt{2mE_F/\hbar^2}$; similarly, in the FM the Fermi wavevectors for the 
majority ($\uparrow$) and minority spin ($\downarrow$) Fermi surfaces are
$k_{F,\uparrow (\downarrow)} = k_F\sqrt{1 +(-) h_{\rm{ex}}}$, where 
$h_{\rm{ex}} = \mu_B|{\bf H}_{\rm{ex}}|/E_F$. The minority Fermi
surface disappears at $h_{\rm{ex}}=1$, and the FM becomes a half-metal.

The gap matrix is
\beq
\hat{\Delta}({\bf r}) \, = \,
\Theta(-x)\,f(\pmb{\nabla})\, i[\psi\hat{\sigma}_{0} + {\bf
    d}\cdot\hat{\pmb{\sigma}}]\, \hat{\sigma}_{y} 
\eeq
where $\psi =\Delta(T)q$ and ${\bf d} = (-i\pmb{\nabla}\times{\bf
  e}_z)\Delta(T)(1-q)/k_F$ are the singlet and
triplet pairing fields, respectively~\cite{frigeriPRL04}. The
parameter $q$ tunes the NCS from 
purely spin-triplet ($q=0$) to purely spin-singlet ($q=1$) pairing. The gap
magnitude $\Delta(T)$ is assumed to have weak-coupling
temperature dependence, with $\Delta(T=0)=\Delta_0$. The  form
factor $f(\pmb{\nabla})$ models different orbital-angular-momentum
pairing states:~$f(\pmb{\nabla}) =1$ for an NCS
with ($s$+$p$)-wave pairing symmetry; and ~$f(\pmb{\nabla}) =
-2\partial_{x}\partial_y / k^2_F$ for a ($d_{xy}$+$p$)-wave pairing state. 
Writing the gap function in momentum space and adopting the 
helicity basis, there is only intra-helicity pairing with distinct  
gaps $\Delta_{{\bf k},\pm} = f(i{\bf k})(\psi_{\bf k} \pm |{\bf d}_{\bf k}|)$
on the spin-split Fermi surfaces. For the circular Fermi surfaces
considered here the negative-helicity gap vanishes at 
$q=q_c$ with $q_c =
k_{F,-}/(k_{F,-}+k_F) \approx 0.55$, which may be regarded as the boundary
between majority-triplet ($q<q_c$) and majority-singlet ($q>q_c$) pairing
states.

\subsection{Scattering wavefunction} \label{subsec:scattering}

The Bogoliubov-de Gennes equation for the
quasiparticle states $\psi({\bf r})$ with energy $E$ is written in Nambu-spin
space as  
\beq
\check{H}({\bf r})\psi({\bf r}) = E\psi({\bf r})\,. \label{eq:BdG}
\eeq
We solve~\eq{eq:BdG} for the scattering states. As an example,
the wavefunction for a $\nu$-helicity electronlike quasiparticle with
wavevector ${\bf k}_{\nu} = (k_{\nu,x},k_y)$ incident upon the FM from the NCS
is given by 
\beqarray
\fl \psi_{e\nu}(k_y,x) = \Theta(-x)e^{ik_yy}
 \bigg(
 \phi^{\rm{NCS}}_{e,{\bf
    k}_{\nu},\nu}e^{ik_{\nu,x}x} \nonumber \\
{}+ \sum_{\nu' =
  \pm}\left\{\left[a^{e}_{\nu\nu'}\phi^{\rm{NCS}}_{h,{\bf k}_{\nu'},\nu'}e^{ik_{\nu',x}x} +
  b^{e}_{\nu\nu'}\phi^{\rm{NCS}}_{e,\widetilde{\bf
      k}_{\nu'},\nu'}e^{-ik_{\nu',x}x}\right]\Theta(k_{F,\nu'}-|k_y|)\right. 
\nonumber \\ 
{} + \left.\left[a^e_{\nu\nu'}\widetilde{\phi}^{\rm{NCS}}_{h,k_y,\nu'}
  e^{\kappa_{\nu'}x} +
  b^e_{\nu\nu'}\widetilde{\phi}^{\rm{NCS}}_{e,k_y,\nu'}e^{\kappa_{\nu'}x}\right]
  \Theta(|k_y|-k_{F,\nu'})\right\} \bigg)
\nonumber \\
{} + \Theta(x)e^{ik_yy}\sum_{\sigma=\uparrow,\downarrow}
\big\{
\left[c^e_{\nu\sigma}\phi^{\rm{FM}}_{h,\sigma}e^{-ik_{\sigma,x}x} 
+ d^e_{\nu\sigma}\phi^{\rm{FM}}_{e,\sigma}e^{ik_{\sigma,x}x}\right]\Theta(k_{F,
  \sigma}-|k_y|)
\nonumber \\
{} + \left[c^e_{\nu\sigma}\phi^{\rm{FM}}_{h,\sigma}e^{-\kappa_{\sigma}x} 
  + d^e_{\nu\sigma}\phi^{\rm{FM}}_{e,\sigma}e^{-\kappa_{\sigma}x}\right]
  \Theta(|k_y|-k_{F,\sigma}) 
  \big\} \,.
\label{eq:psieneg}
\eeqarray
For $x<0$, the ansatz describes Andreev reflection of $\nu'$-helicity holelike 
quasiparticles with wavevector ${\bf k}_{\nu'} = (k_{\nu',x},k_y)$ and
reflection probability 
amplitude $a^e_{\nu\nu'}$ and the normal reflection of $\nu'$-helicity
electronlike quasiparticles with wavevector $\widetilde{\bf k}_{\nu'} =
(-k_{\nu',x},k_{y})$ and 
probability amplitude $b^e_{\nu\nu'}$. For $x>0$, the quasiparticle is
transmitted as a spin-$\sigma$ hole with wavevector $\widetilde{\bf k}_{\sigma} =
(-k_{\sigma,x},k_{y})$ and amplitude $c^e_{\nu\sigma}$ or as a spin-$\sigma$
electron with 
wavevector ${\bf k}_{\sigma}=(k_{\sigma,x},k_y)$ and amplitude
$d^e_{\nu\sigma}$,
respectively. Here we have made the standard assumption of $E\ll E_F$ and
thereby approximate the magnitude of the wavevectors for electrons and holes to be
equal; relaxing this approximation is not expected to qualitatively alter our
results. Note also that the momentum $k_y$ parallel to the interface is a good
quantum number due to translational invariance along the $y$-axis. If $|k_y|$
is larger than the Fermi momentum in a given scattering channel, only
evanescent solutions in this channel are possible. These solutions are
characterized by an 
inverse decay length into the bulk NCS (FM) of $\kappa^{-1}_{\nu}$
($\kappa^{-1}_{\sigma}$).

The wave function~\eq{eq:psieneg} is expressed in terms of the spinors
for the NCS and FM. In the FM, the spinors for electrons (\textit{e}) and holes
(\textit{h}) are 
\numparts
\beqarray
\phi^{\rm{FM}}_{e,\uparrow} =
\frac{1}{\sqrt{2}}\left(e^{-i\eta}\cos\frac{\zeta}{2}, \, \sin\frac{\zeta}{2}, \, 0,
  \, 0\right)^{\!T}  \\
\phi^{\rm{FM}}_{e,\downarrow} =
\frac{1}{\sqrt{2}}\left(-e^{-i\eta}\sin\frac{\zeta}{2}, \, \cos\frac{\zeta}{2}, \, 0,
  \,0\right)^{\!T}  \\
\phi^{\rm{FM}}_{h,\uparrow} =  \frac{1}{\sqrt{2}}\left(0,\,
  0, \, e^{i\eta}\cos\frac{\zeta}{2}, \, \sin\frac{\zeta}{2}\right)^{\!T}  \\
\phi^{\rm{FM}}_{h,\downarrow} =  \frac{1}{\sqrt{2}}\left(0,
  \, 0, \, -e^{i\eta}\sin\frac{\zeta}{2}, \,
  \cos\frac{\zeta}{2}\right)^{\!T}\,.
\eeqarray
\endnumparts
In the NCS, the spinors for electronlike and holelike quasiparticles with
momentum ${\bf k}$ and helicity $\nu$ are given by
\numparts
\beqarray
\phi^{\rm{NCS}}_{e,{\bf k},\nu} = \frac{1}{\sqrt{2}}\left( u_{{\bf k},\nu}, \, -\nu ie^{i\theta}u_{{\bf k},\nu}, \, \nu ie^{i\theta}s_{{\bf k},\nu}v_{{\bf k},\nu}, \, s_{{\bf k},\nu}v_{{\bf
    k},\nu} \right)^{T}  \\
\phi^{\rm{NCS}}_{h,{\bf k},\nu} = \frac{1}{\sqrt{2}}\left(v_{{\bf k},\nu}, \, -\nu ie^{i\theta}v_{{\bf k},\nu}, \, \nu ie^{i\theta}s_{{\bf
    k},\nu}u_{{\bf k},\nu}, \, s_{{\bf k},\nu}u_{{\bf 
    k},\nu} \right)^{T} 
\eeqarray
\endnumparts
where ${\bf k} = k_{F,\nu}(\cos\theta,\sin\theta)$, $s_{{\bf
    k},\nu} = \mbox{sgn}(\Delta_{{\bf k},\nu})$ and 
\numparts
\beqarray
u_{{\bf k},\nu} = \sqrt{\frac{E + \Omega_{{\bf k},\nu}}{2E}}  \\
v_{{\bf k},\nu} = \sqrt{\frac{E - \Omega_{{\bf k},\nu}}{2E}}  \\ 
\Omega_{{\bf k},\nu} = \sqrt{E^2 - \Delta_{{\bf k},\nu}^2}\,.
\eeqarray
\endnumparts
Evanescent solutions in the NCS are characterized by the spinors
\numparts
\beqarray
\widetilde{\phi}^{\rm{NCS}}_{e,k_y,\nu} = \frac{1}{\sqrt{2}}\left(1, \, \nu
\frac{k_y - \kappa_\nu}{k_{F,\nu}}, \, 0, \, 0 \right)^{\!T}  \\
\widetilde{\phi}^{\rm{NCS}}_{h,k_y,\nu} = \frac{1}{\sqrt{2}}\left(0, \, 0, \, 1,
 \, -\nu \frac{k_y + \kappa_\nu}{k_{F,\nu}}\right)^{\!T} 
\eeqarray
\endnumparts
where $\kappa_{\nu} = \sqrt{k_{y}^2-k_{F,\nu}^2}$.

The reflection and transmission amplitudes in~\eq{eq:psieneg} are
determined from the boundary conditions obeyed by the wave function at the
NCS-FM interface.  Firstly, we require that the wavefunction is continuous
at the interface
\beq
\psi_{e\nu}(k_y,x=0^{-}) = \psi_{e\nu}(k_y,x=0^{+}) .
\eeq
To ensure the conservation of probability~\cite{SOCboundaryconditions},
the wavefunction must also obey the condition 
\beq
\fl \left.\partial_{x}\psi_{e\nu}(k_y,x) \right|_{x=0^{+}} -
\left.\partial_{x}\psi_{e\nu}(k_y,x) \right|_{x=0^{-}} =
\hat{\sigma}_0\otimes\left(Zk_F\hat{\sigma}_0 - i\widetilde{\lambda}
k_F \hat{\sigma}_y\right)\psi_{e\nu}(k_y,x=0) 
\eeq
where $Z=Uk_F/E_F$ is a dimensionless constant characterizing the strength
of the insulating barrier. These conditions yield eight coupled equations for
the probability amplitudes.

The calculation for a $\nu$-helicity \emph{holelike} quasiparticle incident on
the interface proceeds analogously. The reflection and transmission
coefficients in this case are denoted by a superscript $h$, i.e.,
$a^{h}_{\nu\nu'}$, $b^{h}_{\nu\nu'}$, etc.

\subsection{The Green's function} \label{subsec:Green}

Generalizing the method of
Refs.~\cite{McMillan68,Ishi70,Furusaki91,KashiwayaTanaka}, we obtain the
following expression for the retarded Green's function
$\check{G}^r_{\rm{NCS}} ({\bf r}, {\bf r}^\prime; E)$ in the NCS as a
$4\times4$ matrix in Nambu-spin space:
\beqarray
\fl \check{G}^r_{\rm{NCS}}({\bf r},{\bf r}';E) =
\int\frac{dk_y}{2\pi}\:\check{G}^r_{\rm{NCS}}(k_y,x,x';E)e^{ik_y(y
  - y')} \nonumber \\
\fl\phantom{\check{G}^r_{\rm{NCS}}({\bf r},{\bf r}';E)} = \int\frac{dk_y}{2\pi}
  \sum_{\nu,\nu'}\frac{m}{i\hbar^2\sqrt{1+\widetilde{\lambda}^2} \, k_{F}
  \cos\theta_{\nu'}}\,
  \frac{E}{\Omega_{k_y,\nu'}}\Theta(k_{F,\nu}-|k_y|)\Theta(k_{F,\nu'}-|k_y|)
\nonumber \\
{}\times \bigg[  \Big\{\delta_{\nu,\nu'}\phi^{\rm{NCS}}_{e,{\bf
      k},\nu}e^{i(k_{\nu,x} + q_\nu)x} \,+\, b^e_{\nu'\nu}\phi^{\rm{NCS}}_{e,\widetilde{\bf
      k},\nu}e^{-i(k_{\nu,x} + q_\nu)x} \nonumber \\ 
{}+ a^e_{\nu'\nu}\phi^{\rm{NCS}}_{h,{\bf
      k},\nu}e^{i(k_{\nu,x} - q_\nu)x}\Big\} \big(\phi^{\rm{NCS}}_{e,{\bf
      k},\nu'}\big)^{\dagger}e^{-i(k_{\nu',x}+ q_{\nu'})x'}   \nonumber \\
{}+ \Big\{\delta_{\nu,\nu'}\phi^{\rm{NCS}}_{h,\widetilde{\bf
      k},\nu}e^{-i(k_{\nu,x} - q_\nu)x} \, + \,b^h_{\nu'\nu}\phi^{\rm{NCS}}_{h,{\bf
      k},\nu}e^{i(k_{\nu,x} - q_\nu)x} \nonumber \\
{}+ a^h_{\nu'\nu}\phi^{\rm{NCS}}_{e,\widetilde{\bf
      k},\nu}e^{-i(k_{\nu,x} +
      q_\nu)x}\Big\}\big(\phi^{\rm{NCS}}_{h,\widetilde{\bf
      k},\nu'}\big)^\dagger e^{i(k_{\nu',x} - q_{\nu'})x'} \bigg]  e^{ik_y(y
      - y')}\,.  \label{eq:NCSGf}
\eeqarray
We present only the result for $x'<x<0$, which is sufficient to obtain all
quantities of interest; the Green's function for $x<x'<0$ has similar form.
In constructing~\eq{eq:NCSGf}, we assume that $|\Delta_{{\bf k},\nu}| =
|\Delta_{\widetilde{\bf k},\nu}| = \Delta_{k_y,\nu}$ (valid for the gaps
considered here), and we 
hence introduce the notation 
$\Omega_{{\bf k},\nu} = \Omega_{k_y,\nu}$. For the Green's function it is
necessary to include the energy-dependent corrections to the electron- and
hole-wavevectors, i.e., $k_{e,\nu,x} \approx k_{\nu,x} + q_{\nu}$ and
$k_{h,\nu,x} \approx k_{\nu,x} - q_{\nu}$, where 
\beq
 q_{\nu} = \frac{m\Omega_{k_y,\nu}}{\hbar^2\sqrt{1+\widetilde{\lambda}^2}\, k_{F}\cos\theta_\nu}\,.
\eeq
Note that we neglect contributions from scattering into evanescent states in
the Green's function ansatz, as enforced by the step functions
in~\eq{eq:NCSGf}.

\subsection{Transverse interface current} \label{subsec:current}

The currents are derived from the continuity equation. The charge-density
operator $\hat{\rho}({\bf r}) =
-e\sum_{\sigma}\hat{\psi}^{\dagger}_{\sigma}({\bf 
  r})\hat{\psi}^{}_{\sigma}({\bf r})$ obeys the Heisenberg equation of motion
\beq
\fl \frac{\partial}{\partial t}\,\hat{\rho}({\bf r}) =
\frac{1}{i\hbar}\, [\hat{\rho}({\bf r}),\hat{H}]
= \frac{1}{i\hbar}\left([\hat{\rho}({\bf r}),\hat{H}_N] + [\hat{\rho}({\bf
    r}),\hat{H}_P]\right)
= -\pmb{\nabla}\cdot\hat{{\bf J}}_{e}({\bf r}) - \pmb{\nabla}\cdot\hat{{\bf
  J}}_{s}({\bf r})  
\eeq
where $\hat{H}_N$ and $\hat{H}_P$ are the normal-state and pairing
Hamiltonians, respectively. The commutators of these Hamiltonians with 
the charge-density operator correspond to the divergence of the so-called
electronic and source current density operators, $\hat{{\bf
J}}_{e}({\bf r})$ and $\hat{{\bf J}}_{s}({\bf r})$,
respectively~\cite{Furusaki91,KashiwayaTanaka}. The total
current density is obtained by calculating the expectation value ${\bf
  J}({\bf r}) = \langle \hat{{\bf J}}_{e}({\bf r}) + 
\hat{{\bf J}}_{s}({\bf r}) \rangle$. Only the electronic term contributes
to the transverse current.

After a lengthy calculation, the $y$-component of the charge current
density in the NCS can be expressed in terms of the retarded Green's function
as 
\beqarray
\fl J_{y}(x) = \frac{1}{\beta}\sum_{i\omega_n}\lim_{{\bf r}'\rightarrow{\bf
    r}}
  \left.{\rm Tr}\left\{\left(\frac{ie\hbar}{4m}\left[\frac{\partial}{\partial y} - \frac{\partial}{\partial
  y'}\right] - \frac{e\lambda}{2 \hbar}\check{S}_x\right)\check{G}^{r}_{\rm{NCS}}({\bf r},{\bf r}';E)\right\}\right|_{E\rightarrow i\omega_n} \nonumber \\
\fl \phantom{J_y(x)} = -\frac{1}{\beta}\sum_{i\omega_n}\int\frac{dk_y}{2\pi}
\left.{\rm Tr}\left\{\left(\frac{e\hbar k_y}{2m} +
\frac{e\lambda}{2 \hbar}\check{S}_x\right)\check{G}^{r}_{\rm{NCS}}(k_y,x,x;E)\right\}\right|_{E\rightarrow
  i\omega_n} \label{eq:curdens} 
\eeqarray
where $\check{S}_x = {\rm diag}(\hat{\sigma}_x,-\hat{\sigma}_x)$. Note that
the first term in the brackets in~\eq{eq:curdens} originates from the
kinetic energy, while the second term is due to the spin-orbit coupling.
The current density can equivalently be written as   
\beqarray
\fl J_y(x) = -\int\frac{dk_y}{2\pi}\int
dE\left\{\frac{e\hbar}{2m}k_y\rho(E,k_y,x) +
\frac{e\lambda}{2 \hbar}\rho^{x}(E,k_y,x)\right\}n_{F}(E) \label{eq:curdensldos} 
\eeqarray
where $n_{F}(E)$ is the Fermi distribution function and 
\beqarray
\rho(E,k_y,x) = -\frac{1}{4\pi}{\rm Im}\left({\rm Tr}\left\{\check{G}_{{\rm NCS}}^r
  (k_y,x,x;E)\right\}\right)
\\
\rho^{x}(E,k_y,x) = -\frac{\hbar}{4 \pi} {\rm Im} \left(
{\rm Tr}\left\{\check{G}_{{\rm NCS}}^r (k_y,x,x; E) \check{S}_{x}\right\}\right)
\eeqarray
are the energy- and momentum-resolved local density of states (LDOS) and
$x$-spin-resolved LDOS at distance $x$ from the
interface, respectively. Although~\eq{eq:curdensldos} is of limited
calculational value, it is useful for interpreting our results.  In
particular, we note that the interface current depends on
the reconstructed electronic structure at the NCS-FM interface
only through these quantities.

From examination of the Green's function, we determine that four distinct
scattering processes contribute to $J_{y}(x)$: intra- and inter-helicity 
normal reflection, and intra- and inter-helicity Andreev reflection. All these
contributions show exponential decay into the bulk NCS on the scale of the
coherence length $\xi_{0} = \hbar v_F/ ( \pi\Delta_0 )$. The interface current
density due to normal reflection and inter-helicity Andreev reflection
processes are, however, further modulated by rapidly oscillating factors with
the length scales ${\sim}(2k_F)^{-1}$ and
${\sim}(2\widetilde{\lambda}k_F)^{-1}$,
respectively. As these length scales are much shorter than $\xi_0$, the
total current contributed by these processes is
negligible and we henceforth ignore them. We thus find the total interface
current $I_y$ in the NCS to be
\beqarray
\fl I_y = \int^{0}_{-\infty} dx\, J_y(x) \nonumber \\
\fl \phantom{I_y} = -\frac{e\hbar}{8\pi m}\sqrt{1 +
  \widetilde{\lambda}^2}k_F\,\frac{1}{\beta}\sum_{i\omega_n}\int \, 
  dk_y  \, k_y 
\sum_{\nu}\frac{1}{k_{F,\nu}}\Theta(k_{F,\nu} - |k_y|)
\nonumber \\
\fl \phantom{I_y = } 
\times
\bigg(\frac{\Delta_{k_y,\nu}}{\Omega^2_{k_y,\nu}}\left(a^{h}_{\nu\nu} +
a^{e}_{\nu\nu}\right)\bigg)
\bigg|_{E\rightarrow i\omega_n} \,. \label{eq:current}
\eeqarray
In deriving~\eq{eq:current}, we find that the
current in the $\nu$-helicity sector contributed by the spin-orbit coupling
is exactly $ \nu\widetilde{\lambda}/k_{F,\nu}$ times that from the kinetic
energy.

A similar expression to~\eq{eq:curdens} can be derived for the interface
charge current density in the FM in terms of the FM Green's function
$\check{G}^r_{\rm{FM}}({\bf r},{\bf r}';E)$. As in the NCS, the Green's
function of the FM includes terms due to normal and Andreev reflection
at the interface. Due to the absence of a pairing potential in the FM region, 
however, the latter processes only appear in the off-diagonal elements of 
the Green's function, and as such they give vanishing contribution to the
trace in the expression for the interface current. Since the remaining
contribution of normal reflection to the charge current density oscillates on
the scale of the inverse Fermi momenta, we expect the integrated current in
the FM to be negligible compared to that in the NCS.

\section{Results} \label{sec:results}

We find that an interface current appears in the NCS for a FM with
magnetization components along the $x$- or $z$-axes, and reverses direction
with the magnetization.  In agreement with Ref.~\cite{Schnyder2013}, there is
  no current for a $y$-polarized FM. In the following we will consider the
two cases of a 
magnetic moment pointing along the positive $x$- and $z$-axes. We present
results only for spin-orbit coupling strength $\widetilde{\lambda}=0.2$ and
 barrier strength $U = 3E_F/k_F$. Note that the symbols in the plots of
the 
currents are included to distinguish the curves and do not  represent
the only data points. In performing the Matsubara sum we utilize a frequency
cutoff of $100\Delta_0$. Our finite-temperature
analysis is extended to zero temperature by replacing the Matsubara summation
in~\eq{eq:current} by an integral over the imaginary frequency axis. 
We express the current in units
of $eE_F/2\pi\hbar$, which is twice the magnitude of the edge current of a
chiral $p$-wave superconductor~\cite{FurMatSig2001,StoRoy2004}. 

\subsection{($s$+$p$)-wave pairing state} \label{subsec:sp}

\begin{figure}
  \begin{center}
  \includegraphics[clip,width=0.7\columnwidth]{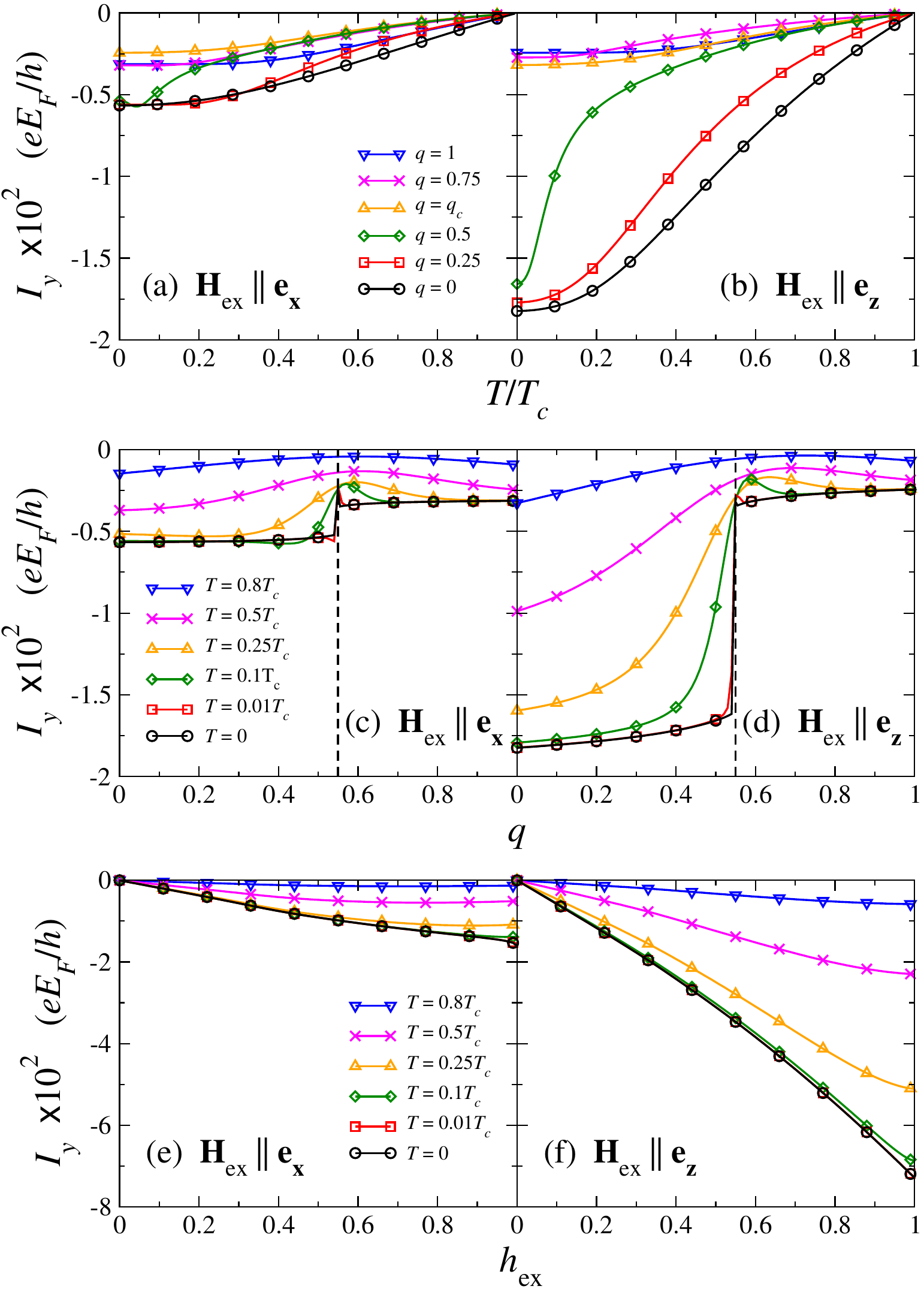}
  \end{center}
  \caption{\label{sp_total} Interface currents in the ($s$+$p$)-wave
    NCS for a FM polarized along the $x$-axis (left column) and $z$-axis
    (right column). Current as a function of: (a), (b) temperature $T$ at
    exchange-field strength $h_{\rm{ex}}=\mu_{B}|{\bf H}_{\rm ex}|/E_F=0.3$;
    (c), (d) singlet-triplet
    parameter $q$ for exchange-field strength $h_{\rm{ex}}=0.3$, where the
    vertical dashed line indicates the critical value $q_c$; (e), (f)
    exchange-field strength $h_{\rm{ex}}$ for singlet-triplet parameter 
    $q=0.25$.}  
\end{figure}

In~\fig{sp_total} we plot the total interface current in the NCS for the
($s$+$p$)-wave pairing state as a function of temperature [panels (a) and (b)],
singlet-triplet parameter $q$ [panels (c) and (d)], and exchange-field strength
$h_{\rm ex}$ [panels (e) and (f)]. For weak to moderate exchange-field strengths
the magnitude of the current in the ($s$+$p$)-wave NCS is always very small
compared to the edge current of a chiral $p$-wave 
superconductor. At low temperatures it converges to a value  
which is weakly dependent on $q$ in both the topologically trivial 
  ($q>q_c$) and nontrivial  ($q<q_c$) states, but has markedly larger
magnitude in the 
latter. Just on the nontrivial side of the topological transition, we see
that the current increases rather steeply at low temperature, note,
e.g., the $q=0.5$ curves in panels (a) and (b).
Plotting the zero-temperature current as a function of $q$
reveals a step discontinuity at $q=q_c$, as seen in panels (c) and
(d). Although the current in the topologically nontrivial NCS is roughly
three times larger for a FM polarized along the 
$z$-axis than along the $x$-axis, in the topologically trivial state the
two magnetization directions give comparable results. The current shows
typical linear-response  behaviour for small exchange-field 
  strengths
$h_{\rm{ex}}$. Approaching the half-metal limit  $h_{\rm ex}=1$, we
observe that the zero-temperature current appears to saturate for the
$x$-polarized FM, but for a magnetization along the $z$-axis the current
increases super-linearly with the exchange field.

\begin{figure}
  \begin{center}
  \includegraphics[clip,width=0.7\columnwidth]{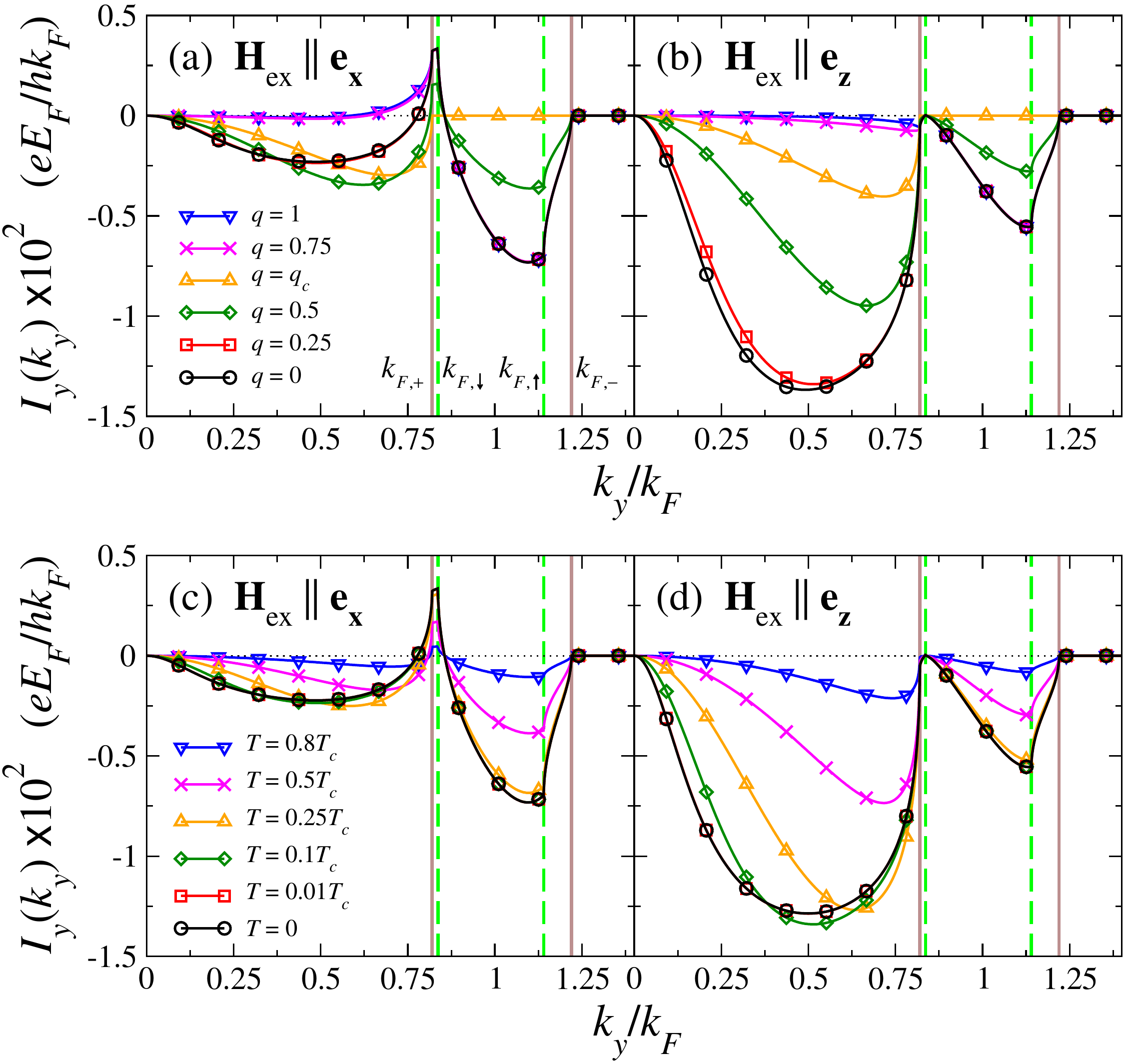}
  \end{center}
  \caption{\label{sp_ky} Momentum-resolved interface current in the
    ($s$+$p$)-wave NCS for a FM polarized along the $x$-axis (left column)
    and $z$-axis (right column) with exchange-field strength $h_{\rm
      ex}=0.3$. (a), (b): The current for various values of the 
    singlet-triplet parameter $q$ at temperature $T=0.1\,T_c$. (c), (d): The
    current for various temperatures $T$ at singlet-triplet parameter
    $q=0.25$. In all panels the vertical brown solid (green 
    dashed) lines are the projections of the positive- and negative-helicity
    Fermi surfaces of the NCS (majority and minority spin Fermi surfaces of
    the FM), as shown in panel (a).  }  
\end{figure}

Insight into the origin of the current  can be gained by examining the 
momentum-resolved quantity $I_{y}(k_y)$, defined as the even part of the 
$k_y$ integrand in~\eq{eq:current}. 
In~\fig{sp_ky} we plot the evolution of $I_y(k_y)$ with the singlet-triplet
parameter $q$ [panels (a), (b)] and the temperature [panels (c), (d)]. This
reveals that the states at $|k_y|<k_{F,+}$ are responsible for the sharp
drop in the current at $q=q_c$, and also for the 
large disparity between the currents for the $x$- and $z$-polarized FM in the 
topologically nontrivial case. In contrast, the momentum-resolved current at
$k_{F,+}<|k_{y}|<k_{F,-}$ is almost independent of the singlet-triplet
parameter for $T=0.1\,T_c$, except  very close to $q=q_c$, where the
negative-helicity gap closes; indeed,
in panels (a) and (b) the curves for $q=1$, $0.75$ and $0.25$ are
  obscured by 
the $q=0$ result. At zero temperature all curves for $q \neq q_c$ are
identical  for $k_{F,+}<|k_{y}|<k_{F,-}$ (not shown). This can be
understood by noting that due to the 
absence of 
positive-helicity states, varying $q$  only
scales the energy dependence of the Andreev reflection coefficients by the
changing negative-helicity gap, which is irrelevant at zero temperature.
On the other hand, the absence of the negative-helicity gap at $q=q_c$
completely suppresses Andreev reflection, and thus there is
vanishing
current. Note also that the current at $k_{F,+}<|k_{y}|<k_{F,-}$ is comparable
for the $x$- and $z$-polarizations of the FM. 

\begin{figure}
  \begin{center}
  \includegraphics[clip,width=0.9\columnwidth]{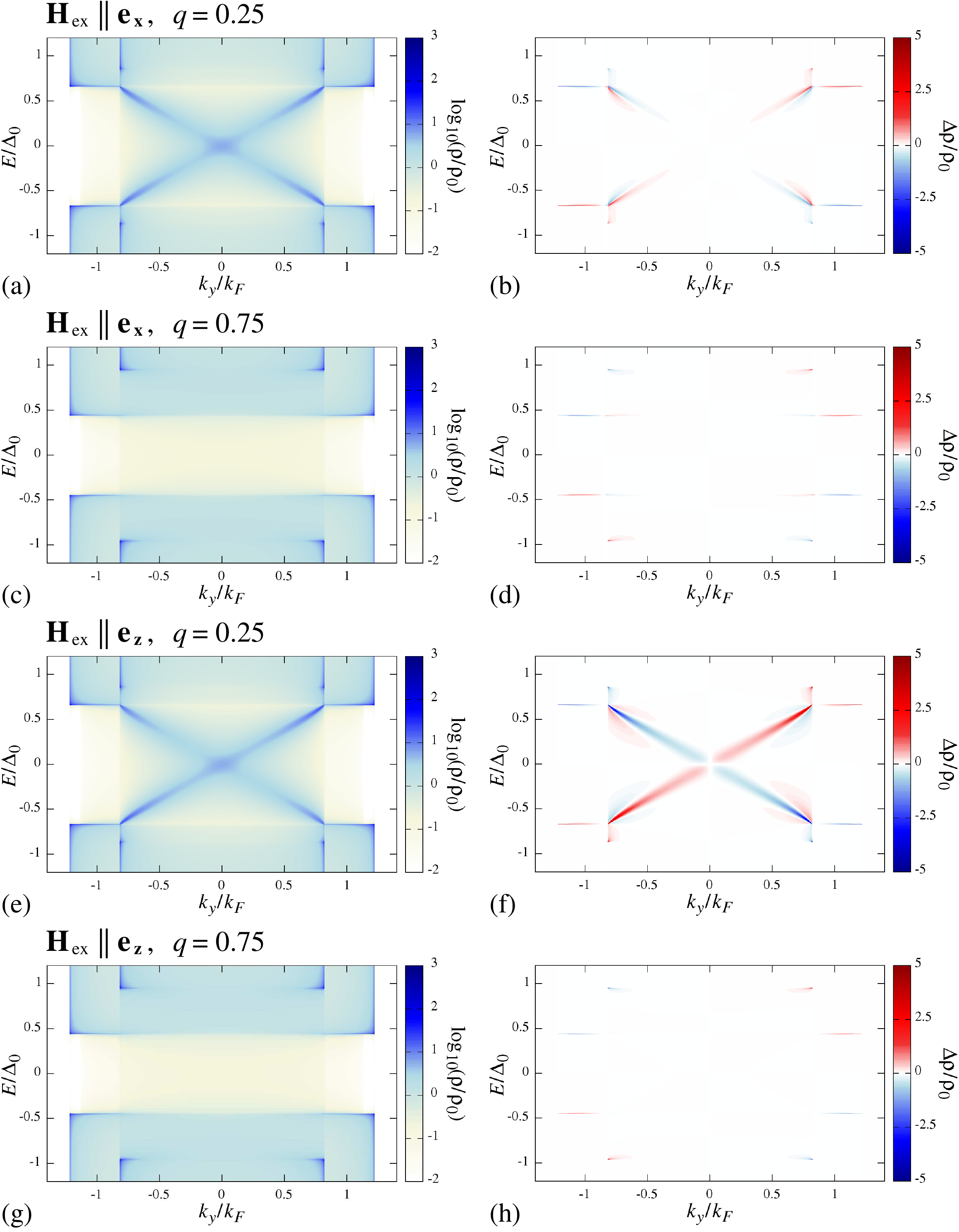}
  \end{center}
  \caption{\label{sp_sdos} Zero-temperature energy- and momentum-resolved
    LDOS in the ($s$+$p$)-wave 
    NCS at the interface with the FM. We present typical results for
    both $x$- and $z$-FM polarizations with $h_{\rm ex}=0.3$, and also both the majority-triplet and
    majority-singlet pairing states. The left column shows the LDOS, while the
    right gives the momentum-antisymmetrized LDOS 
      $\Delta\rho(E,k_y,x) = 
    (\rho(E,k_y,x) - \rho(E,-k_y,x))/2$, both normalized by
    $\rho_0 = m/\pi\hbar^2k_F\sqrt{1 + 
      \widetilde{\lambda}^2}$. We assume an intrinsic broadening of
    $5\times10^{-4}\Delta_0$.}      
\end{figure}

The most interesting features of the current in the ($s$+$p$)-wave NCS are the
zero-temperature step discontinuity at the topological transition ($q=q_c$)
and the much larger current for the majority-triplet 
pairing state. The momentum-resolved results allow us to associate the jump
at $q=q_c$ with a suppression of the current at $|k_y|<k_{F,+}$, which
naturally suggests that the enhanced current in the topologically nontrivial
state is due to the presence of subgap states. This is also
consistent with the sharp low-temperature increase of the current 
near the closing of the negative-helicity gap at $q=q_c$ (see,
e.g., the $q=0.5$
curves in figures~\ref{sp_total}(a) and (b)), as the subgap states can
only contribute to the current at temperatures less than the minimum gap.

To see how the edge states might carry a current, we first examine the
analytically tractable limit of
vanishing spin-orbit coupling and a purely triplet gap. In this case the
Bogoliubov-de Gennes equation~\eq{eq:BdG} in the NCS 
splits up into two $2\times2$ systems with opposite chiral $p$-wave gaps, one
for each Cooper pair $z$-spin orientation $s=\pm 1$. At a vacuum
edge of this
so-called helical superconductor one finds subgap states  with
  dispersion $E_{s=\pm}(k_y) =
s\Delta_0k_y/k_F$~\cite{Qi2009,Iniotakis2007,Vorontsov2008}. When placed next to a  
$z$-polarized metallic FM, the spin-dependent reflectivity ${\cal
  R}_s(k_y)$ leads to an unequal broadening of these
states. In particular, by calculating the poles of the Andreev reflection
coefficients, we obtain  the energy of the subgap states
\beq
E_s(k_y) = \Delta_{0}\sin\left( \arcsin\frac{sk_y}{k_F} + i\ln\sqrt{{\cal
    R}_{s}(k_y)}\right) \,.
\eeq
The left- and right-moving subgap states therefore have finite but different
lifetimes when ${\cal R}_+(k_y)\neq {\cal R}_-(k_y)<1$, and thus
appear as
resonances of different width in the energy- and momentum-resolved LDOS
$\rho(E,k_y,x)$ in the NCS. On the other hand, there is no asymmetry in the
spin-degenerate continuum states, i.e., we find that 
$\rho(E,k_y,x) \neq \rho(E,-k_y,x)$  only for energies
$|E|<\Delta_0$. Furthermore, the integrals of the LDOS at $k_y$ and $-k_y$
over the negative-energy states are unequal $\int^{0}_{-\Delta_0} dE\,
\rho(E,k_y,x) \neq \int^{0}_{-\Delta_0} dE\,
\rho(E,-k_y,x)$, and so from~\eq{eq:curdensldos} we deduce that
there is a finite charge current density in the NCS which is carried
entirely by the subgap states.
We note that the spin-dependent broadening
is not connected to the  total polarization of the states, but rather is
controlled by 
the polarization of the electronlike part of the wavefunction. It follows
that for the helical superconductor the broadening of the  two edge
states is the same  when the FM is polarized along the $x$- or $y$-axis,
and there is hence vanishing interface current.  

It is not possible to rigorously make the above argument in the presence of
spin-orbit coupling or of a singlet gap. We nevertheless
expect that the spin-dependent broadening of the 
edge states should be robust to these complications,  as the spectrum
  evolves continuously as they are switched on. These
spin-mixing terms may also lead to a similar broadening effect for an 
FM polarized along the $x$-axis,  as they lift the degeneracy of the
positive- and negative-helicity states~\cite{LuYip2010}. To test this,
 we plot in~\fig{sp_sdos} the energy- and momentum-resolved LDOS at the
interface for 
systems representing both topologically nontrivial and trivial states and
also both magnetization directions. An asymmetry in
$\rho(E,k_y,x=0^{-})$ is readily visible only for the case of a
majority-triplet
pairing state and a $z$-polarized FM [panel (e)]: the
right-moving subgap state is clearly less broadened than the left-moving state.
Subtle changes in the LDOS are
nevertheless present for the other cases, which can be revealed by plotting
the  momentum-antisymmetrized quantity 
\beqarray
\Delta \rho(E,k_y,x=0^-) = \frac{1}{2}\left[\rho(E,k_y,x=0^-) -
\rho(E,-k_y,x=0^-)\right]  
\eeqarray
in the right column of~\fig{sp_sdos}. The different broadening of the two
subgap states for the $q=0.25$ NCS in contact with a $z$-polarized FM [panel
  (f)] now 
becomes much clearer. On the other hand, the results for the $q=0.25$ NCS in
contact with an $x$-polarized FM [panel (b)]  reveal that the $k_y>0$ and 
$k_y<0$ states have been slightly shifted to higher and lower energies,
respectively. This is consistent with the coupling of their spin polarization
to the exchange field~\cite{Schnyder2013}; a
similar but smaller energy shift for the $z$-polarized FM is masked
in~\fig{sp_sdos}(f) by a much greater broadening. Closer inspection of the
subgap states in~\fig{sp_sdos}(b) nevertheless reveals that the right-moving
state is less broadened than the left-moving state, although the difference in
linewidths is much smaller than for the $z$-polarized FM.
For the majority-singlet cases, the antisymmetrized LDOS reveals that the
changes in the LDOS are  small and restricted to the gap edges. 

This analysis shows that the unequal broadening of the right- and left-moving
edge states is a plausible explanation for the enhanced current in the
topologically 
nontrivial state, and also why there is a much larger current for the
$z$-polarized FM. Furthermore, it
explains why a sharp jump in the current across the topological transition was
not anticipated by the exact-diagonalization calculations presented
in Ref.~\Ref{Schnyder2013}. The broadening of the edge states is dependent on an
imperfect reflectivity, so that spectral weight can leak into the
FM. Clearly this is only possible for a metallic FM; furthermore, in
our ballistic-limit calculation we require that the FM be of width
much larger than the 
NCS's coherence length, so that electron and hole pairs decohere before being
reflected  back at the opposite side of the FM towards the
superconductor. In contrast, in Ref.~\Ref{Schnyder2013} an
insulating FM was considered, which obviously preserves the 
perfect edge reflectivity, and indeed no broadening of the edge
states was found. The  resulting edge current therefore has
completely different characteristics to  that found here, as 
  it is due entirely 
to the induced $x$-spin polarization via the spin-orbit coupling.

\subsection{($d_{xy}$+$p$)-wave pairing state} \label{subsec:dp}

\begin{figure}
  \begin{center}
  \includegraphics[clip,width=0.7\columnwidth]{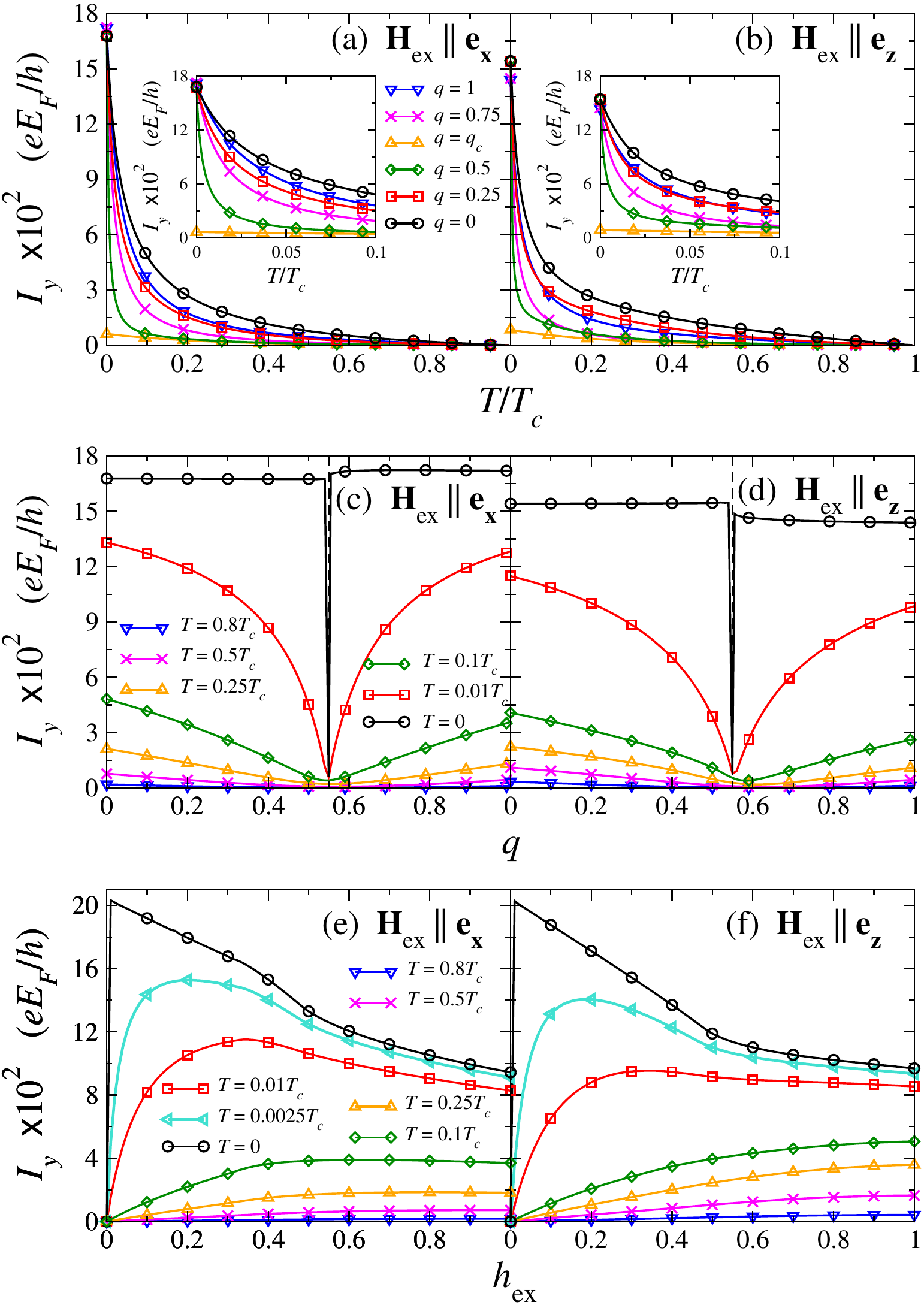}
  \end{center}
  \caption{\label{dp_total} Interface currents in the ($d_{xy}$+$p$)-wave
    NCS for a FM polarized along the $x$-axis (left column) and $z$-axis
    (right column). Current as a
    function of: (a), (b) temperature $T$ for  
    exchange-field strength $h_{\rm{ex}}=0.3$, with insets showing
    low-temperature $T\leq0.1\,T_c$ currents in more detail; (c), (d)
    singlet-triplet parameter $q$ for exchange-field strength
    $h_{\rm{ex}}=0.3$, where the vertical dashed line indicates
    the critical value $q_c$; (e), (f)
    exchange-field strength $h_{\rm{ex}}$ for singlet-triplet parameter
    $q=0.25$. }    
\end{figure}

As shown in~\fig{dp_total}, the currents in the ($d_{xy}$+$p$)-wave NCS are
dramatically different from those in the ($s$+$p$)-wave state. We first note
that 
the sign of the currents is reversed between the two cases, originating from
the additional $\pi$ phase shift acquired by Andreev-reflected
quasiparticles due to the $d_{xy}$ form factor. In further contrast to the
($s$+$p$)-wave case, the currents due to an $x$- and $z$-polarized FM are
quantitatively very similar. More remarkable is the
temperature-dependence of the current [panels (a) and (b)], which  is
characterized by a sharp increase at low temperatures for all $q\neq
q_c$; for $q=q_c$, in contrast, the current grows only slightly below
$T=0.1\,T_c$, see the insets of panels (a) and (b). 
As seen in panels (c) and (d), the magnitude of the zero-temperature
current for $q\neq q_c$ far exceeds that in the ($s$+$p$)-wave case. The
  zero-temperature current is discontinuous at $q_c$: exactly at this point it
  takes a value roughly an order of magnitude smaller than at  
$q=q_{c}\pm 0^{+}$. There also appears to be a small jump in the
current between $q$ on
  either side of $q_c$, accompanied by a 
change of the slope. The difference between the current in the
majority-triplet and majority-singlet regimes is nevertheless much smaller
than for the ($s$+$p$)-wave NCS. At low nonzero temperatures, however, 
the current is sharply suppressed as one approaches $q=q_c$, indicating that
the low temperature enhancement becomes increasingly sharp near the
negative-helicity gap closing,
e.g., see the $q=0.5$ curve in the insets of panels (a) and (b).
Indeed,
the absence of the low-temperature current enhancement for $q=q_c$ implies
that it crucially involves the negative-helicity states. 

The dependence of the current on the exchange field [figures \ref{dp_total}(e)
and
  (f)] shows a remarkable deviation from linear-response behaviour at low
temperatures: for $T\lesssim 0.01\, T_c$ the current grows very rapidly with the
exchange-field strength, before going through a maximum and slowly
decreasing. At zero temperature an infinitesimally small exchange field in the
FM is sufficient to generate a large current in the NCS. We note that there
appears to be a qualitative change in the dependence of the zero-temperature
current on the exchange field at $h_{\rm ex}\approx 0.5$, characterized by a
 change of slope of the low-temperature currents.

\begin{figure}
  \begin{center}
  \includegraphics[clip,width=0.7\columnwidth]{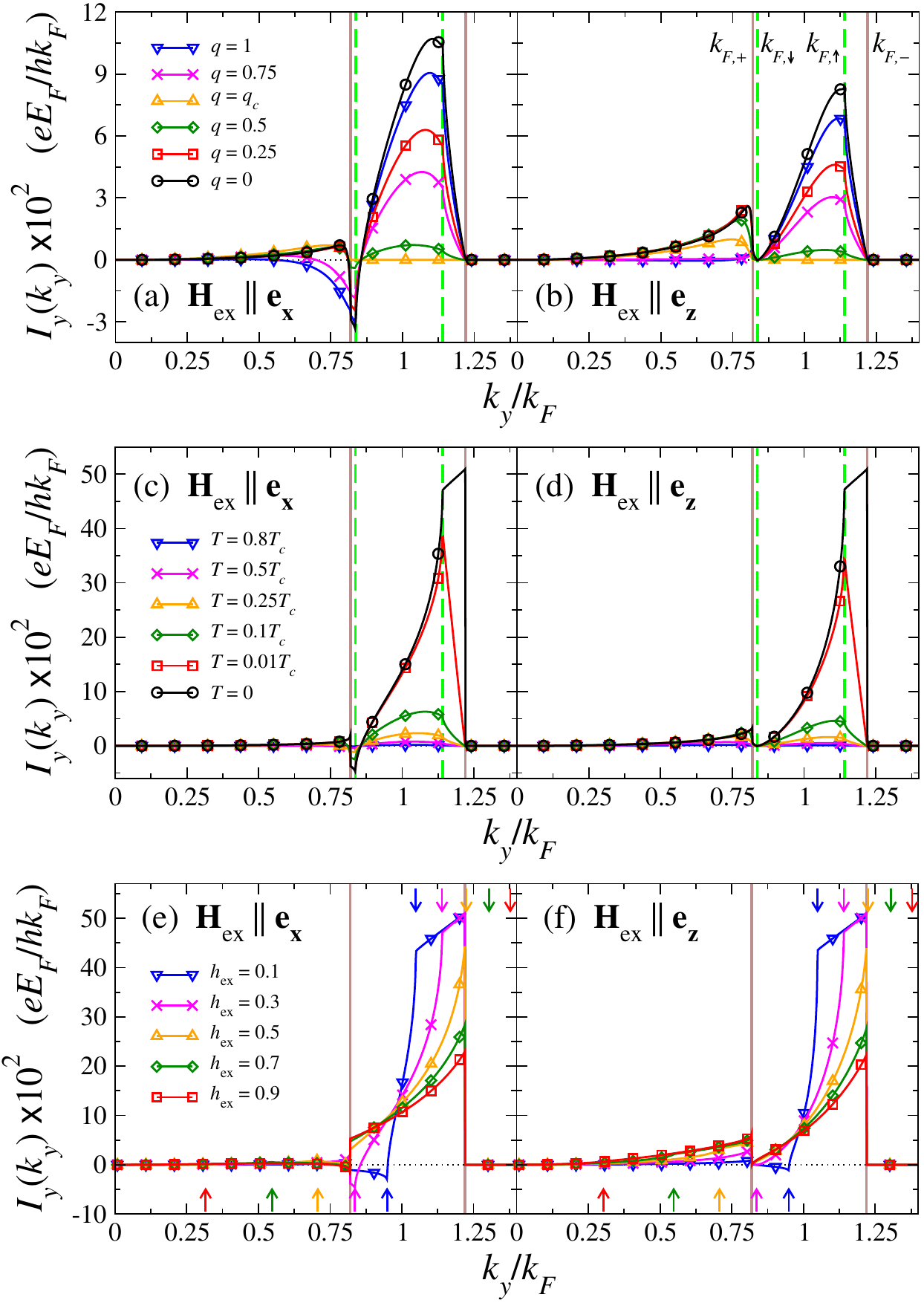}
  \end{center}
  \caption{\label{dp_cky} Momentum-resolved interface current in the
    ($d_{xy}$+$p$)-wave NCS for a FM polarized along the $x$-axis (left column)
    and $z$-axis (right column). We show the momentum-resolved current: (a),
    (b) for various 
    values of the 
    singlet-triplet parameter $q$ at temperature $T=0.1\,T_c$  and exchange
    field
    $h_{\rm ex}=0.3$; (c), (d) for various temperatures $T$ at singlet-triplet
    parameter 
    $q=0.25$ and exchange field $h_{\rm ex}=0.3$; (e), (f) for various exchange fields $h_{\rm ex}$
    at zero temperature and singlet-triplet parameter $q=0.25$. In all panels
    the vertical brown solid lines are the projections 
    of the positive- and negative-helicity 
    Fermi surfaces of the NCS. In (a)-(d) the projected majority and minority
    Fermi surfaces of the FM are given by the vertical green dashed lines,
  while in (e) and (f) they are indicated by the colored arrows. }  
\end{figure}

The momentum-resolved current (\fig{dp_cky}) shows that the low-temperature
enhancement of the current is due entirely to states at
$k_{F,+}<|k_y|<k_{F,-}$, consistent with the critical role of the
negative-helicity gap deduced above. Although there is considerable variation of
this current as a function of $q$ for $T=0.1\,T_c$ [figures \ref{dp_cky}(a) and
(b)], 
at zero temperature there is no variation with $q$ for $q \neq 
q_c$ (not shown), for the same reasons as in the ($s$+$p$)-wave case. Thus,
the change in the total zero-temperature current across the
triplet-singlet boundary [figures \ref{dp_total}(c) and (d)] is due only to
states at
$|k_y|<k_{F,+}$. The temperature-dependence of the $k_{F,+}<|k_y|<k_{F,-}$
current is astonishing: comparing the $T=0.01\,T_c$
and $T=0$ curves in figures \ref{dp_cky}(c) and (d), we observe that while
the current at $|k_y|<k_{F,\uparrow}$ has almost saturated to its
zero-temperature value by $T=0.01\,T_c$, for $k_{F,\uparrow}<|k_y|<k_{F,-}$ the
current more than doubles as the temperature is lowered. For $q=0.25$, the
the zero-temperature current at $k_{F,\uparrow}<|k_y|<k_{F,-}$ accounts for
more than 40\% of the total. Interestingly, in this momentum range the current
displays a linear dependence upon $k_y$, and is almost independent of the
exchange-field strength [figures \ref{dp_cky}(e) and (f)]. The current
at $k_{F,\uparrow}<|k_y|<k_{F,-}$ is therefore clearly somewhat special, and
its disappearance when $k_{F,\uparrow}=k_{F,-}$ at the exchange-field strength
$h_{{\rm ex}}\approx0.49$  closely matches
the qualitative change in the 
exchange-field dependence seen in figures \ref{dp_total}(e) and (f). In closing,
we note that although the momentum-resolved currents at
$|k_y|\gtrsim
k_F$ are very similar for the two polarization orientations, clear
differences are seen at smaller momenta.

\begin{figure}
  \begin{center}
  \includegraphics[clip,width=0.9\columnwidth]{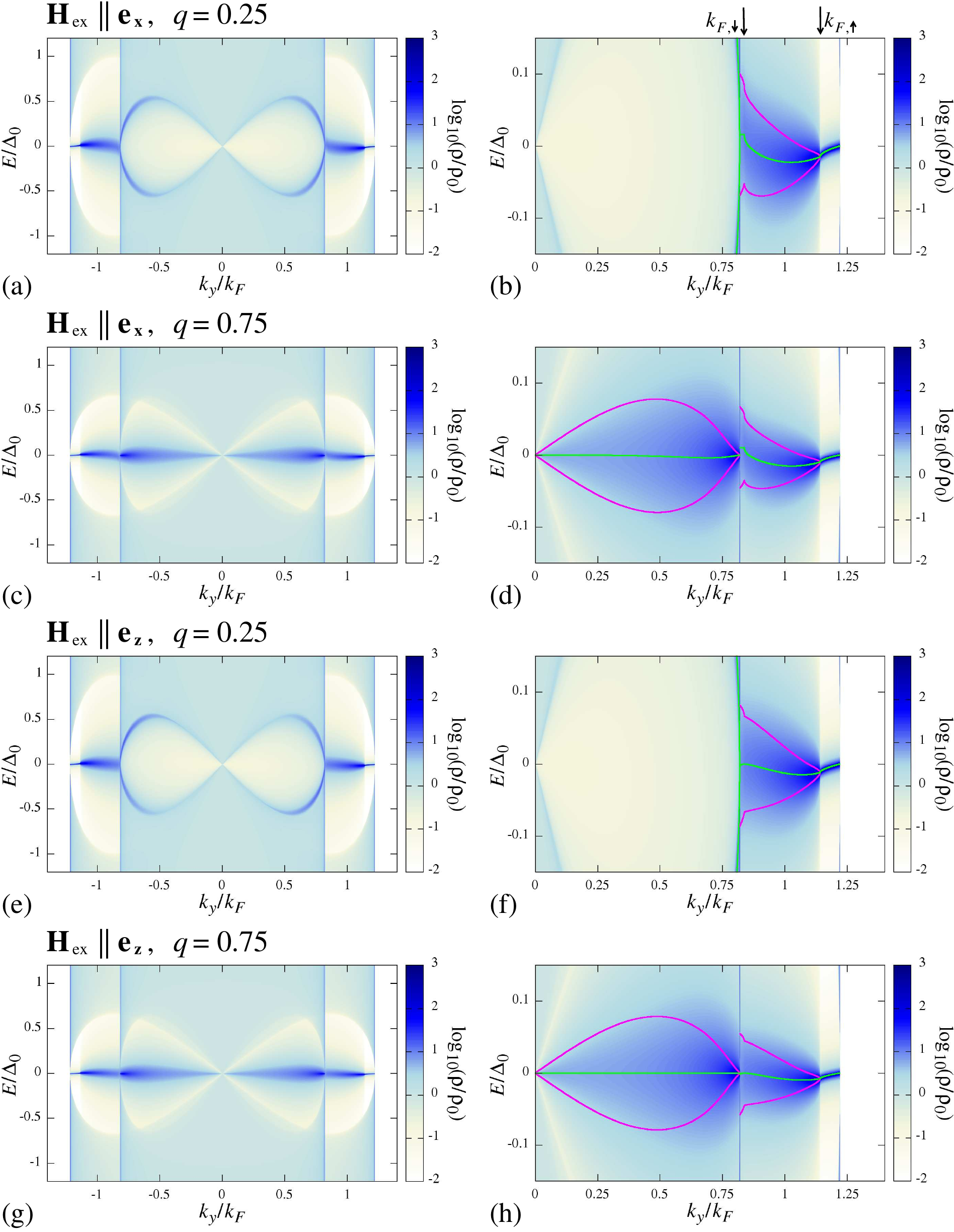}
  \end{center}
  \caption{\label{dp_sdos} Zero-temperature energy- and momentum-resolved 
    LDOS in the ($d_{xy}$+$p$)-wave 
    NCS at the interface with the FM. We present
    typical results for 
    both $x$- and $z$-FM polarizations with $h_{\rm ex}=0.3$, and also both the majority-triplet and
    majority-singlet pairing states.  The left column shows the LDOS for
    momenta $|k_y|<1.4k_F$ and energies $|E|<1.2\Delta_0$,
    while the right column shows a low-energy detail for $k_y>0$. In the
    latter, the light green and dark magenta lines indicate the location of
    the peak in the subgap LDOS and its FWHM, respectively. In panel (b) we
    indicate the projection of the majority- and minority-spin Fermi surfaces
    in the FM. In all panels the
    LDOS is normalized by $\rho_0 = m/\pi\hbar^2k_F\sqrt{1 +
      \widetilde{\lambda}^2}$, and we
    assume an intrinsic broadening $5\times10^{-4}\Delta_0$. }     
\end{figure}

The remarkable temperature dependence of the current in the
($d_{xy}$+$p$)-wave NCS is intimately connected to the coupling of
the exchange field to the topological edge states at
$k_{F,+}<|k_y|<k_{F,-}$. In~Ref.~\Ref{Schnyder2013} it was shown that at a
vacuum edge these states
possess strong $x$- and $z$-spin polarization, with equal magnitude
but opposite sign at
$k_y$ and $-k_y$, as required by time-reversal symmetry. Applying an exchange
field to the edge of the NCS therefore shifts these states in opposite
directions, one above and the other below the Fermi
energy. From~\eq{eq:curdensldos} we see that the first term in the 
zero-temperature momentum-resolved current is proportional to the difference
in the number of states below the Fermi energy at $k_y$ and $-k_y$, while the
second is proportional to the sum of the $x$-spin polarization at these
momenta. The opposite energy shifts of the oppositely polarized flat-band
states at $k_y$ and $-k_y$ thus can contribute a large current. This implies
the breakdown of linear-response behaviour at zero temperature, as then even
an infinitesimal energy shift of these states causes a
discrete change in the number difference and the $x$-spin polarization. 

 For the device studied here, this argument can be directly applied for
$k_{F,\uparrow}<|k_y|<k_{F,-}$ where only evanescent solutions
are possible in the FM and so the subgap states in the NCS have zero
broadening. As seen in the LDOS plots in~\fig{dp_sdos}, these states acquire an
energy shift $\lesssim 0.01\Delta_0$ from the coupling to the exchange
field.  From~\eq{eq:curdensldos}, we
find that the zero-temperature momentum-resolved current for these states has
the form 
\beq
 I_y(k_y) = \frac{eE_F}{4\pi\hbar k_F^2}\,
  \frac{\sqrt{1+\widetilde{\lambda}^2}}{\sqrt{1+\widetilde{\lambda}^2} + \widetilde{\lambda}}\,k_y\, 
\label{eq:bscur}
\eeq
in the $h_{\rm ex}\rightarrow 0$ limit. The induced polarization
at finite values of the exchange field can slightly 
reduce this current, as seen in figures \ref{dp_cky}(e) and (f).  
The momentum-resolved current~\eq{eq:bscur} saturates to
\beq
I_y(k_y) =  \frac{eE_F}{8\pi\hbar k_F^2} k_y\,
\eeq
 as the spin-orbit coupling strength $\widetilde{\lambda}$
diverges. The total zero-temperature current therefore grows linearly with $\widetilde{\lambda}$.
Much larger values of
the low-temperature current than presented here are thus theoretically possible.
Because of the small correction to the energy of the flat-band states, their
contribution to the current only develops at extremely low temperatures
$T<0.01\,T_c$, which is indeed observed in figures \ref{dp_cky}(c) and (d). We
note the
apparent paradox that in spite of the strong spin-polarization of these
  states
at $|k_y|=k_{F,-}$~\cite{Schnyder2013}, the LDOS plots show that the energy
shift vanishes as 
$|k_y|\rightarrow k_{F,-}$, indicating that the states are ``anchored''
to the node of the negative-helicity gap. This is consistent with the
observation that the current saturates at ever
lower temperatures as one approaches $\pm k_{F,-}$ (not shown). 

The flat-band states at $k_{F,+}<|k_y|<k_{F,\uparrow}$ also acquire an energy
shift due to the 
coupling to the exchange field, but the presence of the open scattering
channel in the ferromagnet also gives them a finite lifetime. This  could
not be anticipated by the analysis in~Ref.~\Ref{Schnyder2013}, as discussed 
 in section~\ref{subsec:sp} above. In the right column of~\fig{dp_sdos},
we  mark the maximum in the subgap LDOS by the green lines; the
broadening is quantified by the  
full width at half maximum (FWHM) curves shown in magenta. 
The increasing broadening as $|k_y|\rightarrow k_{F,+}$ results in a
rapid suppression of the zero-temperature momentum-resolved current, as
the imbalance between the integrated weight at $-k_y$ and $k_y$ is
reduced. On the other hand, the broadening allows the current to saturate
at temperatures well above the energy of the subgap maximum. Remarkably, it
still seems 
possible to think of the current in terms of the energy shift of the subgap
state, even when this is much smaller than the broadening. For example, the
zero-crossing of the subgap maximum at $|k_y|\approx k_{F,\downarrow}$ for the
$x$-polarized FM is nicely correlated with a sign change in the
momentum-resolved current [see figures \ref{dp_cky}(a) and (e)], as expected
from
treating the subgap maximum as an unbroadened state. We note that there is no
zero crossing of the subgap maximum for the $z$-polarized FM, and also no sign
change in the current. 

In closing, we note that the LDOS also shows significant structure for states
lying within the projected positive-helicity Fermi surface ($|k_y|<k_{F,+}$),
i.e., broadened dispersing states for majority-triplet pairing, and zero-energy
states for majority-singlet pairing. Although in the latter case there is a
slight shift in the location of the subgap maximum, the large broadening
washes out any contribution to the current, except perhaps for $|k_y|$ very
close to $k_{F,+}$. 

\section{Experimental prospects} \label{sec:experiment}

The interface currents in the NCS-FM junction discussed above contain 
clear signatures of the topological state of the NCS. For the ($s$+$p$)-wave
NCS this is the sharp jump in the current 
at the topological phase transition, while the presence of the
topologically protected flat bands in the ($d_{xy}$+$p$)-wave NCS
directly leads to the sharp increase of the current at low
temperatures  and a nonperturbative dependence on the exchange-field
  strength. The detection of any of these effects would therefore 
be strong evidence for a nontrivial topology of the NCS. 
The  current characteristics of the ($d_{xy}$+$p$)-wave NCS are of
particular interest as  they only arise from nondegenerate flat
bands. In contrast, previous proposals to evidence the flat bands by tunneling
conductance measurements cannot easily distinguish between
nondegenerate and
doubly degenerate states~\cite{Schnyder2012,Tanaka2010,sato2011,Brydon2011}.

The experimental verification of our predictions is nevertheless likely to be
challenging, and must overcome a number of obstacles. 
Foremost is the Meissner effect, as screening in the NCS will exactly
compensate the interface current~\eq{eq:current} for the half-space geometry
considered here. For an NCS of finite width $W$, however, this problem can be
avoided by exploiting the different
length scales of the interface and screening current densities, the coherence
length $\xi_0$ and the penetration depth $\lambda_{L}$,
respectively. The best NCS topological superconductor candidates are extreme
type-II superconductors, e.g., CePt$_3$Si has Ginzburg-Landau parameter
$\kappa = \lambda_{L}/\xi_0 \approx 
140$~\cite{bauerSigristbook}. For such a material it is possible to
choose the sample width 
such that $\xi_0\ll W \ll \lambda_{\rm L}$, which implies that while our
calculation for the edge currents remains valid, screening currents
are very small.
The penetration of the FM's magnetic field into the bulk NCS must then be
considered: although the NCS pairing state is robust to   
weak fields along the $z$-axis, which are more easily kept outside
of the NCS in any case, a field along the $x$-axis can destabilize the
NCS towards a phase where the Cooper pairs acquire a
finite momentum~\cite{AgtKau2007,Loder2012}. Because the consequences of this
for the edge states is unknown, a $z$-polarized FM is therefore a more
favorable choice for experimental study.

The construction of an NCS-FM heterostructure device presents
further difficulties. Firstly, some degree of surface roughness is
unavoidable, which will lead to additional
broadening of the interface states~\cite{Matsumoto1995}.
As long as this does not introduce further energy shifts, however, we
expect that the interface current should persist for weak disorder.
Indeed, we have seen above that even the strongly broadened edge states
of the ($d_{xy}$+$p$)-wave NCS contribute a significant current. 
A more serious problem is the choice of material for the NCS part of the
device. Although there are many examples of bulk NCS, little work has been
done on incorporating them into heterostructures. An alternative
approach is to instead engineer the NCS in the heterostructure, say by
coating a superconducting substrate with a thin normal layer of a material
with strong spin-orbit coupling, so that the former induces a
superconducting gap in the latter. This is in the same spirit as the
well-known proposal to artificially create a topological superconductor
in a quantum wire~\cite{sau2010,Oreg2010},  which has been reported in
recent experiments~\cite{Mourik2012}. All suggestions along these
lines~\cite{sau2010,Oreg2010,fuKanePRL08,Chung2011,PotterLee2012,Takei2013,
Zhang2012} involve the modification of the standard proximity-induced
superconductivity by spin-orbit coupling of the same form present
in a bulk NCS. Using unconventional $d$-wave
cuprate~\cite{Takei2013} or $s_{\pm}$-wave
pnictide~\cite{Zhang2012} superconductors holds particular promise, as
it might then be possible to artificially create the most interesting cases of
($d_{xy}$+$p$)-wave or topologically nontrivial ($s$+$p$)-wave NCS,
respectively. 
No matter how the NCS-FM heterostructure is constructed, however,
there will be some variation of the superconducting gaps close to the 
interface due to the pair-breaking effect of the tunneling barrier and the 
FM~\cite{scferroRMP}. Although our calculation does not account for this,
 the current predicted above should be robust as it
ultimately arises from the spin structure of the bulk condensate.

\section{Summary} \label{sec:summary}

In this paper we have  used a quasiclassical method to study the
properties of the charge
current that appears at the interface between an NCS and a metallic 
FM  in a two-dimensional junction, where each phase is assumed to occupy
a half space. We 
have considered two complementary models of the NCS: a gapped 
($s$+$p$)-wave pairing state, and a gapless ($d_{xy}$+$p$)-wave system. Due to
the contrasting topological structure of the two models, we find
completely different dependences of the interface current on the temperature,
the exchange-field strength $h_{\rm ex}$ and the singlet-triplet parameter
$q$. In both cases we find signatures of the topology in the interface
transport. For the ($s$+$p$)-wave
NCS, the topological transition from the nontrivial to the trivial state is
signaled by a discontinuous drop in the zero-temperature current, due to the
disappearance of the contribution from the subgap states. In the
($d_{xy}$+$p$)-wave NCS, there is an enormous enhancement of the 
current as the temperature approaches zero, and the dependence on
the exchange-field strength becomes singular. This anomalous behaviour
originates from the energy shifts of the spin-polarized flat bands due to the coupling
to the exchange field in the FM. While the results for the
($d_{xy}$+$p$)-wave NCS were anticipated by the analysis of an exchange field
applied directly to the edge~\Ref{Schnyder2013}, the current in the
($s$+$p$)-wave NCS is qualitatively
different due to the broadening of the subgap states by tunneling into the
metallic FM.  We thus find that the mechanism based on flat bands,
relevant for the ($d_{xy}$+$p$)-wave case, is rather robust and independent
of the detailed nature of the FM. Hence, one can
speculate that similar current
characteristics might also be realized in other systems
possessing topologically protected nondegenerate flat bands.

\ack
The authors thank M~Sigrist for useful discussions. APS thanks NORDITA for
its hospitality and financial support. CT acknowledges financial support by
the Deutsche Forschungsgemeinschaft through Research Training Group GRK 1621.

\end{document}